\documentclass[useAMS]{mn2e}

\usepackage{times}
\usepackage{epsfig}

\title[Mode coupling in radially stratified discs] {Linear coupling
of modes in 2D radially stratified astrophysical discs}
\author[Tevzadze et al.]{A. G. Tevzadze$^1$, G. D.
Chagelishvili$^{1,2}$, G. Bodo$^3$ and P. Rossi$^3$ \\
$^1$ Georgian National Astrophysical Observatory, Chavchavadze
State University, Tbilisi, Georgia \\
$^2$ Nodia Institute of Geophysics, Georgian Academy of Sciences,
Tbilisi, Georgia \\
$^3$ INAF -- Osservatorio Astronomico di Torino, strada
dell'Osservatorio 20, I-10025 Pino Torinese, Italy}

\begin{document}

\date{\today}

\maketitle

\begin{abstract}
We investigate mode coupling in a two dimensional compressible disc
with radial stratification and differential rotation. We employ the
global radial scaling of linear perturbations and study the linear
modes in the local shearing sheet approximation. We employ a
three-mode formalism and study the vorticity (W), entropy (S) and
compressional (P) modes and their coupling properties. The system
exhibits asymmetric three-mode coupling: these include mutual
coupling of S and P-modes, S and W-modes, and asymmetric coupling
between the W and P-modes. P-mode perturbations are able to generate
potential vorticity through indirect three-mode coupling. This
process indicates that compressional perturbations can lead to the
development of vortical structures and influence the dynamics of
radially stratified hydrodynamic accretion and protoplanetary discs.
\end{abstract}

\section{Introduction}

The recent increased interest in the analysis of hydrodynamic disc
flows is motivated, on one hand, by the study of turbulent
processes, and, on the other,  by the investigation of regular
structure formation in protoplanetary discs. Indeed, many
astrophysical discs are thought to be neutral or having ionization
rates too low to effectively couple with magnetic field. Among these
are cool and dense areas of protoplanetary discs, discs around young
stars, X-ray transient and dwarf nova systems in quiescence (see
e.g. Gammie and Menou 1998, Sano et al. 2000, Fromang, Terquem and
Balbus 2002). Observational data shows that astrophysical discs
often exhibit radial gradients of thermodynamic variables (see e.g.
Sandin et al. 2008, Issela et al. 2007). To what extent these
inhomogeneities affect the processes occurring in the disc is still
subject open to investigations. It has been found that strong local
entropy gradients in the radial direction may drive the Rossby wave
instability (Lovelace et al. 1999, Li et al. 2000) that transfers
thermal to kinetic energy and leads to vortex formation. However, in
astrophysical discs, radial stratification is more likely weak. In
this case, the radial entropy (temperature) variation on the global
scale leads to the existence of baroclinic perturbations over the
barotropic equilibrium state. This more appropriate situation has
recently become a subject of extensive study.

Klahr and Bodenheimer (2003) pointed out that the radial
stratification in the disc can lead to the global baroclinic
instability. Numerical results show that the resulting state is
highly chaotic and transports angular momentum outwards. Later Klahr
(2004) performed a local 2D linear stability analysis of a radially
stratified flow with constant surface density and showed that
baroclinic perturbations can grow transiently during a limited time
interval. Johnson and Gammie (2005) derived analytic solutions for
3D linear perturbations in a radially stratified discs in the
Boussinesq approximation. They find that leading and trailing waves
are characterized by  positive and negative angular momentum flux,
respectively. Later Johnson and Gammie (2006) performed numerical
simulations, in the local shearing sheet model, to test the radial
convective stability and the effects of baroclinic perturbations.
They found no substantial instability due to the radial
stratification. This result reveals a controversy over the issue of
baroclinic instability. Presently,  it seems that nonlinear
baroclinic instability is an unlikely development in the local
dynamics of sub-Keplerian discs with weak radial stratification.

Potential vorticity production, and the formation and development of
vortices in radially stratified discs have been studied by Petersen
et al. (2007a,b) by using pseudospectral simulations in the
anelastic approximation.  They show that the existence of thermal
perturbations in the radially stratified disc flows leads to the
formation of vortices. Moreover, stronger vortices appear in discs
with higher temperature perturbations or in simulations with higher
Reynolds numbers, and the transport of angular momentum may be both
outward and inward.

Keplerian differential rotation in the disc is characterized by a
strong velocity shear in the radial direction. It is known that
shear flows are non-normal and exhibit a number of transient
phenomena due to the non-orthogonal nature of the operators (see
e.g. Trefethen et al.  1993). In fact, the studies described above
did not take into account the possibility of mode coupling and
energy transfer between different modes due to the shear flow
induced mode conversion.  Mode coupling is inherent to shear flows
(cf.  Chagelishvili et al.  1995) and often, in many respects,
defines the role of perturbation modes in the system dynamics and
the further development of nonlinear processes. Thus, a correct
understanding of the energy exchange channels between different
modes in the linear regime is vital for a correct understanding of
the nonlinear phenomena.

Indications of the shear induced mode conversion can be found in a
number of previous studies. Barranco and Marcus 2005 report that
vortices are able to excite inertial gravity waves during 3D
spectral simulations. Brandenburg and Dintrans (2006) have studied
the linear dynamics of perturbation SFH to analyze nonaxisymetric
stability in the shearing sheet approximation. Temporal evolution of
the perturbation gain factors reveal a wave nature after the radial
wavenumber changes sign. Compressible waves are present, along with
vortical perturbations, in the simulation by Johnson \& Gammie
(2005b) but their origin is not particularly discussed.

In parallel, there are a number of papers that focus on the
investigation of the shear induced mode coupling phenomena. The
study of the linear coupling of modes in Keplerian flows has been
conducted in the local shearing sheet approximation (Tevzadze et al.
2003,2008) as well as in 2D global numerical simulations (Bodo et
al. 2005, hereafter B05). Tevzadze et al. (2003) studied the linear
dynamics of three-dimensional small scale perturbations (with
characteristic scales much less then the disc thickness) in
vertically (stably) stratified Keplerian discs. They show, that
vortex and internal gravity wave modes are coupled efficiently. B05
performed global numerical simulations of the linear dynamics of
initially imposed two-dimensional pure vortex mode perturbations in
compressible Keplerian discs with constant background pressure and
density. The two modes possible in this system are effectively
coupled: vortex mode perturbations are able to generate
density-spiral waves. The coupling is, however, strongly asymmetric:
the coupling is effective for wave generation by vortices, but not
viceversa. The resulting dynamical picture points out the importance
of  mode coupling and the necessity of considering compressibility
effects for processes with characteristic scales of the order or
larger than the disc thickness. Bodo et al. (2007) extended this
work to nonlinear amplitudes and found that mode coupling is an
efficient channel for energy exchange and is not an artifact of the
linear analysis. B05 is particularly relevant to the present study,
since it studies the dynamics of mode coupling in 2D unstratified
flows and is a good starting point for a further extension to
radially stratified flows. Later, Heinemann \& Papaloizou (2009a)
derived WKBJ solutions of the generated waves and performed
numerical simulations of the wave excitation by turbulent
fluctuations (Heinemann \& Papaloizou 2009b).

In the present paper we study the linear dynamics of perturbations
and analyze shear flow induced mode coupling in the local shearing
sheet approximation. We investigate the properties of mode coupling
using qualitative analysis within the three-mode approximation.
Within this approximation we tentatively distinguish vorticity,
entropy and pressure modes. Quantitative results on mode conversion
are derived numerically. It seems that a weak radial
stratifications, while being a weak factor for the disc stability,
still provides an additional degree of freedom (an active entropy
mode), opening new options for velocity shear induced mode
conversion, that may be important for the system behavior. One of
the direct result of mode conversion is the possibility of linear
generation of the vortex mode (i.e., potential vorticity) by
compressible perturbations. We want to stress the possibility of the
coupling between  high and low frequency perturbations, considering
that high frequency oscillations have been often neglected during
previous investigations in particular for protoplanetary discs.

Conventionally there are two distinct viewpoints commonly employed
during the investigation of hydrodynamic astrophysical discs. In one
case (self gravitating galactic discs) the emphasis is placed on the
investigation of the dynamics of spiral-density waves and vortices,
although normally present in numerical simulations, are thought to
play a minor role in the overall dynamics. In the other case
(non-self-gravitating hydrodynamic discs) the focus is on the
potential vorticity perturbations and density-spiral waves are often
thought to play a minor role. Here, discussing the possible (multi)
mode couplings, we want to draw attention to the possible flaws of
these simplified views (see e.g. Mamatsashvili \& Chagelishvili
2007). In many cases,  mode coupling makes different perturbation to
equally participate in the dynamical processes despite of a
significant difference in their temporal scales.

In the next section we present mathematical formalism of our study.
We describe three mode formalism and give schematic picture of the
linear mode coupling in the radially sheared and stratified flow.
Numerical analysis of the mode coupling is presented in Sec. 3. We
evaluate mode coupling efficiencies at different radial
stratification scales of the equilibrium pressure and entropy. The
paper is summarized in Sec. 4.

\section{Basic equations}

The governing ideal hydrodynamic equations of a  two-dimensional,
compressible disc flows in polar coordinates are:
\begin{equation}
{\partial \Sigma \over \partial t}  + {1 \over r} {\partial \left( r
\Sigma V_r \right) \over
\partial r}  + {1 \over r} {\partial \left( \Sigma V_\phi \right)\over
\partial \phi}  = 0~,~~~~~~~~~~~~~~~~~~~~~~
\end{equation}
\begin{equation}
{\partial  V_r \over \partial t}  +  V_r{\partial  V_r \over
\partial r}+ {V_\phi \over r}{\partial V_r \over
\partial \phi}  - {V_\phi^2 \over r} = -{1 \over \Sigma}
{\partial P \over \partial r}  - {\partial \psi_g \over \partial r}
~,~~~~~~
\end{equation}
\begin{equation}
{\partial  V_\phi \over \partial t}  +  V_r{\partial  V_\phi \over
\partial r}+ {V_\phi \over r}{\partial V_\phi \over
\partial \phi} +
{V_r V_\phi \over r} = -{1 \over \Sigma r}{\partial P\over
\partial \phi}~,~~~~~~~~~
\end{equation}
\begin{equation}
{\partial  P \over \partial t}  +  V_r{\partial  P \over
\partial r}+ {V_\phi \over r}{\partial P \over
\partial \phi}
= - {\gamma P} \left( {1 \over r} {\partial (r V_r) \over
\partial r}  + {1 \over r} {\partial V_\phi \over \partial
\phi}  \right)~,
\end{equation}
where  $V_r$ and $V_\phi$ are the flow radial and azimuthal
velocities respectively. $P(r,\phi)$, $\Sigma(r,\phi)$ and $\gamma~$
are respectively the pressure, the surface density and the adiabatic
index. $\psi_g$ is the gravitational potential of the central mass,
in the absence of self-gravitation $~(\psi_g \sim -{1 / r})$. This
potential  determines the Keplerian angular velocity:
\begin{equation}
{\partial \psi_g \over \partial r} = \Omega_{Kep}^2 r ~,~~~~
\Omega_{Kep} \sim r^{-3/2};
\end{equation}

\subsection{Equilibrium state}

We consider an axisymmetric $(\partial / \partial \phi \equiv 0),~$
azimuthal $(\bar {V}_{r} = 0)~$ and differentially rotating  basic
flow: $\bar {V}_{\phi}= \Omega(r)r$. In the 2D radially stratified
equilibrium (see Klahr 2004), all variables are assumed to follow a
simple power law behavior:
\begin{equation}
\bar {\Sigma}(r) = \Sigma_0 \left( {r \over r_0}
\right)^{-\beta_\Sigma},~~~~\bar {P}(r) = P_0\left( {r \over r_0}
\right)^{-\beta_P} ~,
\end{equation}
where overbars denote equilibrium and $\Sigma_0$ and $P_0$ are the
values of the equilibrium surface density and pressure at some
fiducial radius $r = r_0$. The entropy can be calculated as:
\begin{equation}
\bar S  =  \bar P \bar \Sigma^{-\gamma} = - \left(r \over r_0
\right)^{-\beta_S} ~,
\end{equation}
where
\begin{equation}
\beta_S \equiv \beta_P - \gamma \beta_\Sigma ~.
\end{equation}
$S$ is sometimes called potential temperature, while the physical
entropy can be derived as $\bar S = C_V \log S + \bar S_0$.

This equilibrium shows a deviation from the Keplerian profile due to
the radial stratification:
$$
\Delta \Omega^2(r) = \Omega^2(r) - \Omega^2_{Kep}= {1 \over r {\bar
{\Sigma}(r)}} {\partial {\bar{P}(r)}\over \partial r} =
~~~~~~~~~~~~~~~~~~~~~~~~~~~~~~~~~~~~~~~~~~~~~~~~~~~~~~~~~~~~~~~~~~~~
$$
\begin{equation}
~~~~~~~~~~~~~~~~~~~~~~~~~~~~~~~~~~~~~~~~~~~~~~~~~~~~~~~~~~ = - {P_0
\over \Sigma_0} {\beta_P \over r_0^2}\left( {r \over r_0}
\right)^{\beta_\Sigma-\beta_P-2} ~.
\end{equation}
Hence, the described state is sub-Keplerian or super-Keplerian when
the radial gradient of pressure is negative ($\beta_P>0$) or
positive ($\beta_P<0$), respectively. Although these discs are
non-Keplerian, they are still rotationally supported, since the
deviation from the Keplerian profile is small: $~\Delta
\Omega^2(r)\ll \Omega^2_{Kep}$.

\subsection{Linear perturbations}

We split the physical variables into mean and perturbed parts:
\begin{equation}
\Sigma(r,\phi) = {\bar {\Sigma}(r)} + {{\Sigma}^\prime(r,\phi)} ~,
\end{equation}
\begin{equation}
P(r,\phi) = {\bar{P}(r)} + P^\prime(r,\phi) ~,
\end{equation}
\begin{equation}
V_r(r,\phi) =  V_r^\prime(r,\phi) ~,
\end{equation}
\begin{equation}
V_\phi(r,\phi) = \Omega(r) r + V_\phi^\prime(r,\phi) ~.
\end{equation}
In order to remove background trends from the perturbations we
employ the global radial power law scaling for perturbed quantities:
\begin{equation}
\hat \Sigma(r)  \equiv \left({r \over r_0}\right)^{-\delta_\Sigma}
\Sigma^\prime(r) ~,
\end{equation}
\begin{equation}
\hat P(r) \equiv \left({r \over r_0}\right)^{-\delta_P} P^\prime(r)
~,
\end{equation}
\begin{equation}
\hat {\bf V}(r)  \equiv \left({r \over r_0}\right)^{-\delta_V} {\bf
V}^\prime(r) ~.
\end{equation}

After the definitions one can get the following dynamical equations
for the scaled perturbed variables:
\begin{equation}
\left\{ {\partial \over \partial t} + \Omega(r) {\partial \over
\partial \phi} \right\} {\hat \Sigma \over \Sigma_0} +
~~~~~~~~~~~~~~~~~~~~~~~~~~~~~~~~~~~~~~~~~~~~~~~~~~~~~~~~~~~~~~~
\end{equation}
$$
\left( {r \over r_0} \right)^{-\beta_\Sigma-\delta_\Sigma+\delta_V}
\left[ {\partial \hat V_r \over \partial r} + {1 \over r} {\partial
\hat V_\phi \over
\partial \phi} + {1+\delta_V-\beta_\Sigma \over r} \hat V_r \right]
= 0 ~,
$$
\begin{equation}
\left\{ {\partial \over \partial t} + \Omega(r) {\partial \over
\partial \phi} \right\} \hat V_r - 2\Omega(r) \hat V_\phi +
\end{equation}
$$
{c_s^2 \over \gamma} \left({r \over r_0}
\right)^{\beta_\Sigma+\delta_P-\delta_V} {\partial \over \partial r}
{\hat P \over P_0} + c_s^2 {\delta_P \over \gamma r_0} \left( {r
\over r_0} \right)^{\beta_\Sigma+\delta_P-\delta_V-1} {\hat P \over
P_0} +
$$
$$
~~~~~~~~~~~~~~~~~~~~~~~~~~~~~~ c_s^2 {\beta_P \over \gamma r_0}
\left( {r \over r_0}
\right)^{2\beta_\Sigma+\delta_\Sigma-\beta_P-\delta_V-1} {\hat
\Sigma \over \Sigma_0} = 0 ~,
$$
\begin{equation}
\left\{ {\partial \over \partial t} + \Omega(r) {\partial \over
\partial \phi} \right\} \hat V_\phi + \left( 2 \Omega(r) +
r {\partial \Omega(r) \over \partial r} \right) \hat V_r +
~~~~~~~~~~~~~~~
\end{equation}
$$
~~~~~~~~~~~~~~~~~~~~~~~~~~~~~~~~~~~~~~ {c_s^2 \over \gamma r_0}
\left( {r \over r_0} \right)^{\beta_\Sigma+\delta_P-\delta_V-1}
{\partial \over \partial \phi} {\hat P \over P_0} = 0 ~,
$$
\begin{equation}
\left\{ {\partial \over \partial t} + \Omega(r) {\partial \over
\partial \phi} \right\} {\hat P \over P_0} +
\end{equation}
$$
\gamma \left( {r \over r_0} \right)^{-\beta_P+\delta_V-\delta_P}
\left[ {\partial \hat V_r \over
\partial r} + {1 \over r} {\partial \hat V_\phi \over \partial
\phi} + {1+\delta_V-\beta_P/\gamma \over r} \hat V_r \right] = 0 ~,
$$
where $c_s^2 = \gamma P_0/\Sigma_0$ is the squared sound speed at
$r=r_0$.

\subsection{Local approximation}

The linear dynamics of perturbations in differentially rotating
flows can be effectively analyzed in the local co-rotating shearing
sheet frame (e. g., Goldreich \& Lynden-Bell 1965; Goldreich \&
Tremaine 1978). This approximation simplifies the mathematical
description of flows with inhomogeneous velocity. In the radially
stratified flows the spatial inhomogeneity of the governing
equations comes not only from the equilibrium velocity, but from the
pressure, density and entropy profiles as well. In this case we
first re-scale perturbations in global frame in order to remove
background trends from linear perturbations, rather then use
complete form of perturbations to the equilibrium (see Eqs. 14-16).
Hence, using the re-scaled linear perturbation ($\hat P$, $\hat
\Sigma$, $\hat {\rm \bf V}$) we may simplify local shearing sheet
description as follows. Introduction of a local Cartesian
co-ordinate system:
\begin{equation}
x \equiv r - r_0~,~~~~ y \equiv r_0 (\phi - \Omega_0 t)~,~~~~{x
\over r_0} ,~ {y \over r_0} \ll 1~,
\end{equation}
\begin{equation}
{\partial \over \partial x} = {\partial \over \partial r}~,~~~
{\partial \over \partial y} = {1 \over r_0}{\partial \over
\partial \phi}~,~~~ {\partial \over \partial t} =
{\partial \over \partial t} -  r_0 \Omega_0 {\partial \over
\partial y},
\end{equation}
where $\Omega_0$ is the local rotation angular velocity at $r=r_0$,
transforms global differential rotation into a local radial shear
flow and the two Oort constants define the local shear rate:
\begin{equation}
A \equiv {1 \over 2} r_0 \left[ {\partial \Omega(r) \over \partial
r}\right]_{r=r_0}~,~~~~~~~~~~~~~~~~~~~~~~~~~~~~~
\end{equation}
\begin{equation}
B \equiv - {1 \over 2} \left[ r{\partial \Omega(r)\over \partial r}
+ 2\Omega(r) \right]_{r=r_0}= -A - \Omega_0~.
\end{equation}
Hence, the equations describing the linear dynamics of perturbations
in local approximation read as follows:
\begin{equation}
\left\{ {\partial \over \partial t} + 2Ax {\partial \over
\partial y} \right\} {\hat P \over \gamma P_0} +
\end{equation}
$$
~~~~~~~~~~~~~~~~~~~~~~~~~~~~~ \left[ {\partial \hat V_x \over
\partial x} + {\partial \hat V_y \over \partial y} +
{1+\delta_V-\beta_P/\gamma \over r_0} \hat V_x \right] = 0 ~,
$$
\begin{equation}
\left\{ {\partial \over \partial t} + 2Ax {\partial \over
\partial y} \right\} {\hat V_x} - 2\Omega_0 \hat V_y +
\end{equation}
$$
~~~~~~~~~~~~~~~~~~~~~ c_s^2 \left[ {\partial \over \partial x} {\hat
P \over \gamma P_0} + {\delta_P + \beta_P/\gamma \over r_0} {\hat P
\over \gamma P_0} - {\beta_P \over \gamma r_0} {\hat S \over \gamma
P_0}\right] = 0 ~,
$$
\begin{equation}
\left\{ {\partial \over \partial t} + 2Ax {\partial \over
\partial y} \right\} {\hat V_y} - 2B \hat V_y + c_s^2
{\partial \over \partial y} {\hat P \over \gamma P_0} =0 ~,
\end{equation}
\begin{equation}
\left\{ {\partial \over \partial t} + 2Ax {\partial \over
\partial y} \right\} {\hat S \over \gamma P_0} - {\beta_S
\over \gamma r_0} \hat V_x = 0 ~,
\end{equation}
where $\hat S $ is the entropy perturbation:
\begin{equation}
\hat S \equiv \hat P - c_s^2 \hat \Sigma ~.
\end{equation}
Now we may adjust the global scaling law of perturbations in order
to simplify the local shearing sheet description (see Eqs. 25,26):
\begin{equation}
1 + \delta_V - \beta_P/\gamma = 0 ~,
\end{equation}
\begin{equation}
\delta_P + \beta_P/\gamma = 0 ~.
\end{equation}

Let us introduce spatial Fourier harmonics (SFHs) of perturbations
with time dependent phases:
\begin{equation}
\left( \begin{array}{c} {\hat V}_x({\bf r},t) \\ {\hat V}_y({\bf r},t) \\
{{\hat P}({\bf r},t) / \gamma P_0} \\ {{\hat S}({\bf r},t) / \gamma
P_0} \end{array} \right) =
\left( \begin{array}{r} u_x({\bf k}(t),t) \\ u_y({\bf k}(t),t) \\
-{\rm i} p({\bf k}(t),t) \\ s({\bf k}(t),t)
\end{array} \right) \times
\end{equation}
$$
~~~~~~~~~~~~~~~~~~~~~~~~~~~~~~~~~~~~~~~~~~~~~~~~~~~~~~~~~~~~~~~~~
\exp \left( {\rm i} k_x(t) x + {\rm i} k_y y \right) ~,
$$
with
\begin{equation}
k_x(t) = k_x(0) - 2Ak_y t~.
\end{equation}

Using the above expansion and Eqs. (27-30), we obtain a compact ODE
system that governs the local dynamics of SFHs of perturbations:
\begin{equation}
{{\rm d} \over {\rm d} t} p - k_x(t) u_x - k_y u_y = 0 ~,
\end{equation}
\begin{equation}
{{\rm d} \over {\rm d} t} u_x - 2 \Omega_0 u_y + c_s^2 k_x(t) p -
c_s^2 k_P s = 0 ~,
\end{equation}
\begin{equation}
{{\rm d} \over {\rm d} t} u_y - 2 B u_x + c_s^2 k_y p = 0 ~,
\end{equation}
\begin{equation}
{{\rm d} \over {\rm d} t} s - k_S u_x = 0 ~.
\end{equation}
where
\begin{equation}
k_P = {\beta_P \over \gamma r_0} ~~~ k_S = {\beta_S \over \gamma
r_0} ~.
\end{equation}
The potential vorticity:
\begin{equation}
W \equiv k_x(t)u_y - k_y u_x - 2B p ~,
\end{equation}
is a conserved quantity in barotropic flows: $W = {\rm const.}$ when
$k_P=0$.

\subsection{Perturbations at rigid rotation}

The dispersion equation of our system can be obtained in the
shearless limit ($A=0$, $B=-\Omega$). Hence, using Fourier expansion
of perturbations in time $\propto \exp({\rm i} \omega t)$, in the
shearless limit, we obtain:
\begin{equation}
\omega^4 - \left( c_s^2 k^2 + 4 \Omega_0^2 - c_s^2 \eta \right)
\omega^2 - c_s^4 \eta k_y^2 = 0 ~,
\end{equation}
where
\begin{equation}
\eta \equiv k_P k_S = {\beta_P \beta_S \over \gamma^2 r_0^2} ~.
\end{equation}
Solutions of the Eq. (40) describe a compressible density-spiral
mode and a convective mode that involves perturbations of entropy
and potential vorticity. For weakly stratified discs $(\eta \ll
k^2)$, we find the frequencies are:
\begin{equation}
\bar \omega_{p}^2 = c_s^2 k^2 + 4 \Omega_0^2 ~,
\end{equation}
\begin{equation}
\bar \omega_{c}^2 = - {c_s^4 \eta k_y^2 \over c_s^2 k^2 + 4
\Omega_0^2} ~.
\end{equation}
High frequency solutions ($\bar \omega_{p}^2$) describe the
density-spiral waves and will be referred later as the P-modes. Low
frequency solutions ($\bar \omega_{c}^2$), instead, describe radial
buoyancy mode due to the stratification. In barotropic flows
($\eta=0$) this mode is degenerated into stationary zero frequency
vortical solution. Therefore, we may refer to it as a baroclinic
mode. The mode describes instability when $\eta>0$. In this case the
equilibrium pressure and entropy gradients point in the same
direction. Klahr (2004) has anticipated such result, although worked
in the constant surface density limit ($\beta_\Sigma=0$). The same
behavior has been obtained for axisymmetric perturbations in Johnson
and Gammie (2005). For comparison, in our model baroclinic
perturbations are intrinsically non-axisymmetric. Hence, our result
obtained in the rigidly rotating limit shows that the local
exponential stability of the radial baroclinic mode is defined by
the Schwarzschild-Ledoux criterion:
\begin{equation}
{{\rm d} \bar P \over {\rm d} r} {{\rm d} \bar S \over {\rm d} r}
> 0 ~.
\end{equation}

The dynamics of linear modes can be described using the modal
equations for the eigenfunctions:
\begin{equation}
\left\{ {{\rm d}^2 \over {\rm d} t^2} + \bar \omega_{p,c}^2 \right\}
\Phi_{p,c}(t) = 0 ~,
\end{equation}
where $\Phi_p(t)$ and $\Phi_c(t)$ are the eigenfunctions of the
pressure and convective (baroclinic) modes, respectively. The form
of these functions can be derived from Eqs. (34-41) in the shearless
limit:
\begin{equation}
\Phi_{p,c}(t) = (\bar \omega_{p,c}^2+c_s^2 \eta) p(t) - 2 \Omega_0
W(t) - c_s^2 k_P k_x s(t) ~.
\end{equation}
All physical variables in our system ($p$, $u_x$, $u_y$, $s$) can be
expressed by the two modal eigenfunctions and their first time
derivatives ($\Phi_{p,c}$, $\Phi_{p,c}^\prime$). Hence, we can fully
derive the perturbation field of a specific mode individually by
setting the eigenfunction of the other mode equal to zero.

As we will see later, the Keplerian shear leads to the degeneracy of
the convective buoyancy mode. In this case only the shear modified
density-spiral wave mode eigenfunction can be employed in the
analysis.

\subsection{Perturbations in shear flow: mode coupling}

It is well known that velocity shear introduces non-normality into
the governing equations that significantly affects the dynamics of
different perturbations. In this case we benefit from the shearing
sheet transformation and seek the solutions in the form of the
so-called Kelvin modes. These originate from the vortical solutions
derived in seminal paper by Kelvin (1887). In fact, as it was argued
lately (see e.g., Volponi and Yoshida 2002), the shearing sheet
transformation leads to some sort of generalized modal approach.
Shear modes arising in such description differ from linear modes
with exponential time dependence in many respects. Primarily, phases
of these continuous spectrum shear modes vary in time through
shearing wavenumber; their amplitudes can be time dependent; and
most importantly, they can couple in limited time intervals. On the
other hand, shear modes can be well separated asymptotically, where
analytic WKBJ solution for the each mode can be increasingly
accurate. In the following, we will simply refer to these shearing
sheet solution as ``modes''.

The character of shear flow effects significantly depend on the
value of velocity shear parameter. To estimate the time-scales of
the processes we compare the characteristic frequencies of the
linear modes $|\bar \omega_p|$, $|\bar \omega_c|$ and the velocity
shear $|A|$. In order to speak about the modification of the linear
mode by the velocity shear, the basic frequency of the mode should
be higher than the one set by shear itself: $\omega^2 > A^2$.
Otherwise the modal solution can not be used to calculate
perturbation dynamics, since perturbations will obey the shear
induced variations at shorter timescales.

In quasi-Keplerian differentially rotating discs with weak radial
stratification:
\begin{equation}
\bar \omega_p^2 \gg A^2 ~~~~ {\rm and}  ~~~ \bar \omega_{c}^2 \ll
A^2 ~, {~~~ \rm when ~~~} {\beta_P \beta_S \over \gamma^2} \ll 1 ~.
\end{equation}
In this case the convective mode diverges from its modal behavior
and is strongly affected by the velocity shear: the thermal and
kinematic parts obey shear driven dynamics individually. Therefore,
we tentatively distinguish shear driven vorticity (W) and entropy
(S) modes. On the contrary, the high frequency pressure mode is only
modified by the action of the background shear. Hence, we assume the
above described three mode (S, W and P) formalism as the framework
for our further study.

For the description of the P mode in differential rotation we define
the function:
\begin{equation}
\Psi_p(t) = \omega_p^2(t) p(t) - 2\Omega_0 W(t) - c_s^2 k_P k_x(t)
s(t) ~,
\end{equation}
where
\begin{equation}
\omega_p^2(t) = c_s^2 k^2(t) - 4B \Omega_0 ~.
\end{equation}
This can be considered as the generalization of the $\Phi_P(t)$
eigenfunction for the case of the shear flow, by accounting for the
temporal variation of the radial wavenumber.

In order to analyze the mode coupling in the considered limit, we
rewrite Eqs. (34-39) as follows:
\begin{equation}
\left\{ {{\rm d}^2 \over {\rm d} t^2} + f_p {{\rm d} \over {\rm d}
t} + \omega_p^2 - \Delta \omega_p^2 \right\} \Psi_p = \chi_{pw} W +
\chi_{ps} s ~,
\end{equation}
\begin{equation}
\left\{ {{\rm d} \over {\rm d} t} + f_s \right\} s = \chi_{s p 1}
{{\rm d}  \Psi_p \over {\rm d} t} + \chi_{s p 2} \Psi_p + \chi_{s w}
W ~,
\end{equation}
\begin{equation}
{{\rm d} W \over {\rm d} t}  = \chi_{ws} s ~,
\end{equation}
where $f_p$ and $\Delta \omega_p^2$ describe the shear flow induced
modification to the P-mode
\begin{equation}
f_p =  4 A { k_x k_y \over k^2}  - 2 {(\omega_p^2)^\prime \over
\omega_p^2 } ~,
\end{equation}
\begin{equation}
\Delta \omega_p^2 =  {(\omega_p^2)^{\prime \prime} \over \omega_p^2}
+ f_p {(\omega_p^2)^\prime \over \omega_p^2 } + 8AB{k_y^2 \over k^2}
~,
\end{equation}
parameter $f_s$ describes the modification to the entropy mode
\begin{equation}
f_s = c_s^2 \eta {k_x^2 (\omega_p^2)^\prime \over k^2 \omega_p^4} ~,
\end{equation}
and $\chi$ parameters describe the coupling between the different
modes:
\begin{equation}
\chi_{pw} =  2 \Omega_0 \Delta \omega_p^2(t) + 4A {k_y^2 \over k^2}
\omega_p^2 ~,
\end{equation}
\begin{equation}
\chi_{ps} = c_s^2 k_P k_x \left( \Delta \omega_p^2 + 4B {k_y \over
k_x} {(\omega_p^2)^\prime \over \omega_p^2} - 8AB {k_y^2 \over k^2}
\right) ~,
\end{equation}
\begin{equation}
\chi_{s p 1} = {k_S k_x \over k^2 \omega_p^2 }~,
\end{equation}
\begin{equation}
\chi_{s p 2} = -{k_S k_x \over k^2 \omega_p^2 } \left(
{(\omega_p^2)^\prime \over \omega_p^2} + 2B {k_y \over k_x} \right)
~,
\end{equation}
\begin{equation}
\chi_{s w} = -{2 \Omega k_S k_x \over k^2 \omega_p^2} \left(
{(\omega_p^2)^\prime \over \omega_p^2} + 2B {k_y \over k_x} + {k_y
\omega_p^2 \over 2 \Omega k_x } \right) ~,
\end{equation}
\begin{equation}
\chi_{w s} = -c_s^2 k_P k_y ~.
\end{equation}
Here prime denotes temporal derivative.

Equations (50-52) describe the linear dynamics of modes and their
coupling in the considered three mode model. In this limit, our
interpretation is that the homogeneous parts of the equations
describe the individual dynamics of modes, while the right hand side
terms act as a source terms and describe the mode coupling. This
tentative separation is already fruitful in a qualitative
description of mode coupling.

Dynamics of the density-spiral wave mode in the differential
rotation can be described by the homogeneous part of the Eq. (50).
Homogeneous part of Eq. (51) describes the modifications to the
entropy dynamics. Inhomogeneous parts of the Eqs. (50-52) reveal
coupling terms between the three linear modes that originate due to
the background velocity shear and radial stratification. We analyze
the mode coupling dynamics numerically, but use the coupling $\chi$
coefficients for qualitative description.

\begin{figure}
\begin{center}
\includegraphics[width=80mm]{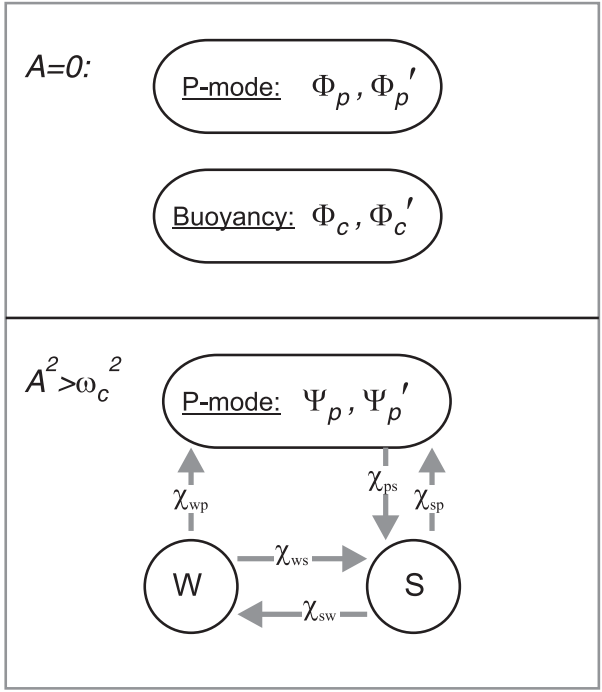}
\end{center}
\caption{ Mode coupling scheme. In the zero shear limit two second
order modes P-mode and buoyancy mode with eigenfunctions $\Phi_p$
and $\Phi_c$ are uncoupled. In the shear flow, when the
characteristic time of shearing is shorter then the buoyancy mode
temporal variation scale ($A^2 > \bar \omega_c^2$), we use three
mode formalism. In this limit we consider the coupling of the P, W,
and S modes. $\chi$ parameters describe the strength of the coupling
channel. Asymmetry of the mode coupling is revealed in the fact that
compressible oscillations of the pressure mode are not able to
directly generate potential vorticity, but still do so via
interaction with S-mode and farther baroclinic ties with W-mode.}
\label{coupling}
\end{figure}

The sketch illustration of the mode coupling in the above described
three-mode approximation can be seen in Fig. \ref{coupling}. The
figure reveals a complex picture of the three mode coupling that
originates by the combined action of velocity shear and radial
stratification.

\begin{figure}
\begin{center}
\includegraphics[width=80mm]{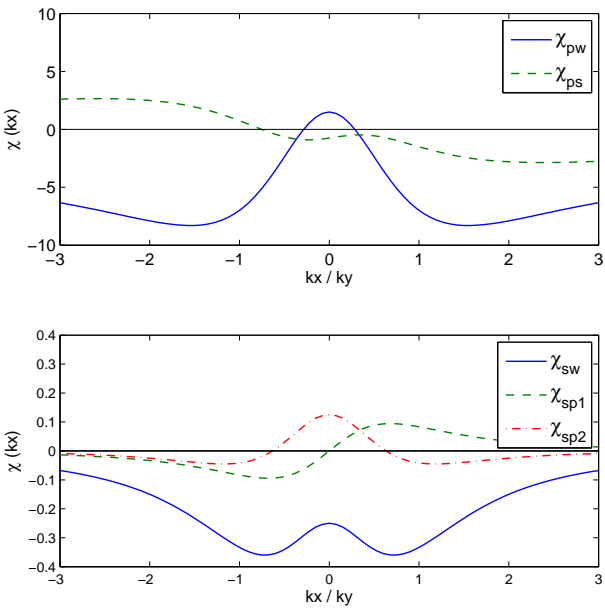}
\end{center}
\caption{The coupling $\chi$ parameters vs. the ratio of radial to
azimuthal wavenumbers $k_x(t)/k_y$ when latter passes through zero
value during the time interval $\Delta t = 4 \Omega_0^{-1}$. Here
$k_y = H^{-1}$, $k_P = k_S = 0.5 H^{-1}$.  } \label{chi}
\end{figure}

The temporal variation of the coupling coefficients during the swing
of the perturbation SFHs from leading to trailing phases is shown in
Fig. \ref{chi}.  The relative amplitudes of the $\chi_{pw}$ and
$\chi_{ps}$ parameters reveal that potential vorticity is a somewhat
more effective source of P mode perturbations when compared to the
entropy mode. On the other hand, it seems that S mode excitation
sources due to potential vorticity ($\chi_{sw}$) can be stronger
when compared with the P-mode sources ($\chi_{sp1}$, $\chi_{sp2}$).

The effect of the stratification parameters on the mode coupling is
somewhat more apparent. First, we may conclude that the excitation
of the entropy mode, which depends on the parapeters $\chi_{sp1}$,
$\chi_{sp2}$ and $\chi_{sw}$ is generally a stronger process for
higher entropy stratification scales $k_S$ (see Eqs. 58-60). Second,
we see that the generation of the potential vorticity depending on
the $\chi_{ws}$ parameter proceeds more effectively at hight
pressure stratification scale $k_P$. And third, we see profound
asymmetry in the three-mode coupling: P-mode is not coupled with the
W-mode {\it directly}.

A quantitative estimate of the mode excitation parameters can be
done using a numerical analysis. In this case, the amplitudes of the
generated W and S modes can be estimated through the values of
potential vorticity or entropy outside the coupling area. In order
to quantify the second order P mode dynamics we define its modal
energy as follows:
\begin{equation}
E_P(t) \equiv |\Psi_p(t)^\prime|^2 + \omega_p(t)^2 |\Psi_P(t)|^2 ~.
\end{equation}
This quadratic form is a good approximation to the P mode energy in
the areas where it obeys adiabatic dynamics: $k_x(t)/k_y \gg 1$.

The presented qualitative analysis suggests that perturbations of the
density-spiral waves can generate entropy perturbations not only due
to the flow viscosity (not included in our formalism), but also
kinematically, due to the velocity shear induced mode coupling. The
generated entropy perturbations should further excite potential
vorticity through baroclinic coupling. Hence, it seems that in
baroclinic flows, contrary to the barotropic case, P-mode
perturbations are able to generate potential vorticity through a
three-mode coupling mechanism: P $\to$ S $\to$ W. We believe that
traces of the described mode coupling can be also seen in Klahr
2004, where the process has not been fully resolved due to the
numerical filters used to remove higher frequency oscillations.

\section{Numerical Results}

In order to study the mode coupling dynamics in more detail we
employ numerical solutions of  Eqs. (34-37). We impose initial
conditions that correspond to the one of the three modes and use a
standard Runge-Kutta scheme for numerical integration (MATLAB ode34
RK implementation). Perturbations corresponding to the individual
modes at the initial point in time are derived in the Appendix A.

\subsection{W-mode: direct coupling with S and P-modes}

\begin{figure}
\begin{center}
\includegraphics[width=84mm]{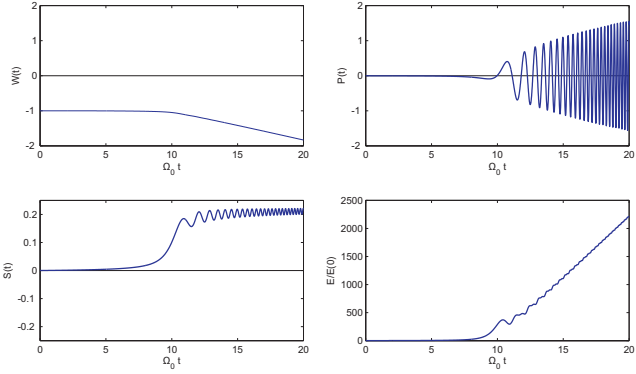}
\end{center}
\caption{Evolution of the W-mode SFH in the flow with
$k_x(t)=-30H^{-1}$, $k_y=2H^{-1}$ and equilibrium with growing
baroclinic perturbations $k_P=k_S=0.2H^{-1}$. Mode coupling occurs
in the vicinity of $t=10 \Omega_0^{-1}$, where $k_x(t)=0$.
Excitation of the P and S-modes are clearly seen in the panels for
pressure ($P$) and entropy ($S$) perturbations. Perturbations of the
potential vorticity start to grow due to the baroclinic coupling
with entropy perturbations. } \label{SFH_w1}
\end{figure}

In this subsection we consider the dynamics of SFH when only the
perturbations of potential vorticity are imposed initially. As it is
known from previous studies (see Chagelishvili et al. 1997, Bodo et
al. 2005) vorticity perturbations are able to excite acoustic modes
nonadiabatically in the vicinity of the area where $k_x(t)=0$. Here
we observe a similar, but more complex, behavior  of mode coupling.
The W-mode is able to generate P and S-modes simultaneously. Fig.
\ref{SFH_w1} shows the evolution of the W-mode perturbations in a
flow with growing baroclinic perturbations ($\eta>0$). The results
show the excitation of both S and P-mode perturbations due to mode
coupling that occurs in a short period of time in the vicinity of
$t=10$. The following growth of the negative potential vorticity is
due to the baroclinic coupling of entropy and potential vorticity
perturbations.

\begin{figure}
\begin{center}
\includegraphics[width=84mm]{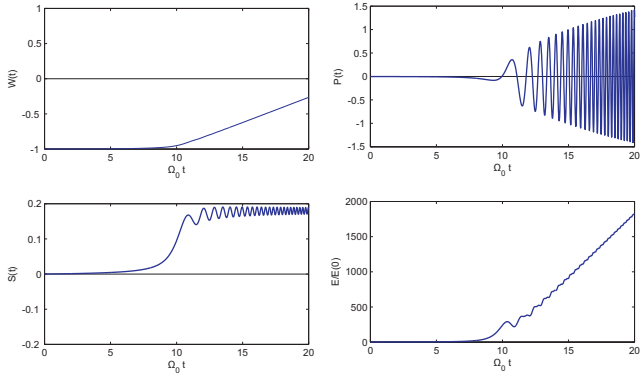}
\end{center}
\caption{Evolution of the W-mode SFH in the flow with
$k_x(t)=-30H^{-1}$, $k_y=2H^{-1}$ and equilibrium with positive
$\eta$: $k_P=-0.2H^{-1}$, $k_S=0.2H^{-1}$. Interestingly, SFH
dynamics shows the decay of potential vorticity after the mode
coupling and excitation of S- and W-modes at $t=10 \Omega_0^{-1}$.
The latter fact is normally anticipated process in the flows that
are baroclinically stable. } \label{SFH_w2}
\end{figure}

Fig. \ref{SFH_w2} shows the evolution of potential vorticity SFH in
flows with negative $\eta$. After the mode coupling and generation
of P and S-modes, we observe a decrease of potential vorticity. This
represents the well known fact that stable stratification (positive
Richardson number) can play a role of ``baroclinic viscosity'' on
the vorticity perturbations.

Numerical calculations show that the efficiency of the mode coupling
generally decreases as we increase  the azimuthal wavenumber $k_y$
corresponding to an increase of the density-spiral wave frequency:
lower frequency waves couple more efficiently.

To test the effect of background stratification parameters on the
mode coupling, we calculate the amplitude of the entropy and the
energy of the P-mode perturbations generated in  flows with
different pressure and entropy stratification scales. The amplitudes
are calculated after a $10 \Omega_0^{-1}$ time interval from the
change in sign of the radial wave-number.  In this case, modes are
well isolated and the energy of the P mode can be well defined.

\begin{figure}
\begin{center}
\includegraphics[width=84mm]{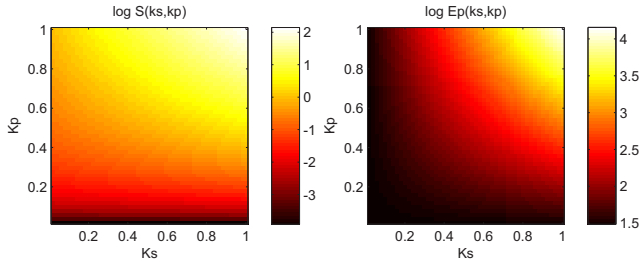}
\end{center}
\caption{Surface graph of the generated S and P-mode amplitudes at
$ky=2H^{-1}$, $kx=-60H^{-1}$, and different values of $k_P$ and
$k_S$. Initial perturbations are normalized to set E(0)=1.
Excitation amplitudes of the entropy perturbations show stronger
dependence of the $k_S$ (left panel), while both entropy and
pressure scales are important (approximately $k_S k_P$ dependence)
for the generation of P-modes (right panel). See electronic edition
of the journal for color images.} \label{surf_w}
\end{figure}

Fig. \ref{surf_w} shows the results of these calculations. It seems
that the mode coupling efficiency is higher with stronger radial
gradients. In particular, numerical results generally verify our
qualitative results that the S-mode generation predominantly depends
on the entropy stratification scale $k_S$. Therefore, P-mode
excitation is stronger for higher values of $\eta$.

\subsection{S-mode: direct coupling with W and P-modes}

\begin{figure}
\begin{center}
\includegraphics[width=84mm]{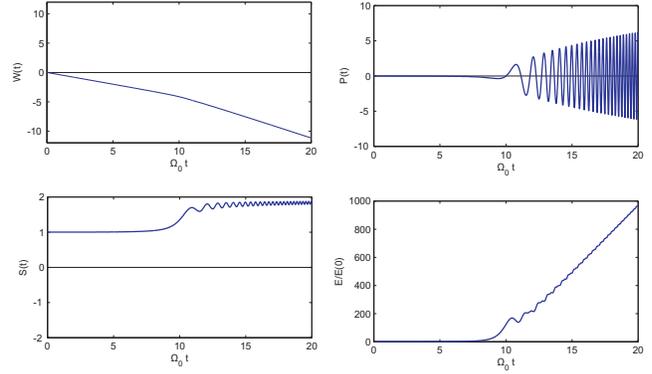}
\end{center}
\caption{Evolution of the S-mode SFH in the flow with
$k_x(t)=-30H^{-1}$, $k_y=2H^{-1}$ and equilibrium with growing
baroclinic perturbations $k_P=k_S=0.2H^{-1}$. Perturbations of the
potential vorticity are coupled grow from the beginning due to the
baroclinic coupling with entropy perturbations. Excitation of the
P-mode is clearly seen in the panel for pressure ($P$), while the
panel for entropy perturbations ($S$) shows swing amplification in
the nonadiabatic area around $k_x(t)=0$. Change of the amplitude of
the entropy SFH affects the growth factor of potential vorticity
SFH.} \label{SFH_s1}
\end{figure}

Fig. \ref{SFH_s1} shows the evolution of the S-mode SFH in a flow
with growing baroclinic perturbations.  Here we observe two shear
flow phenomena: mode coupling and transient amplification. Entering
the nonadiabatic area (around $t = 10$) the entropy SFH is able to
generate the P-mode, while undergoing transient amplification
itself. The transient growth of entropy is unsubstantial and the
growth rate decreases with the growth of $k_y$. The W-mode is
instead constantly coupled to entropy perturbations through
baroclinic forces, although higher entropy perturbations at later
times give an  higher rate of growth of potential vorticity. The
total energy of perturbations is however dominated, at the end, by
the P mode.

\begin{figure}
\begin{center}
\includegraphics[width=84mm]{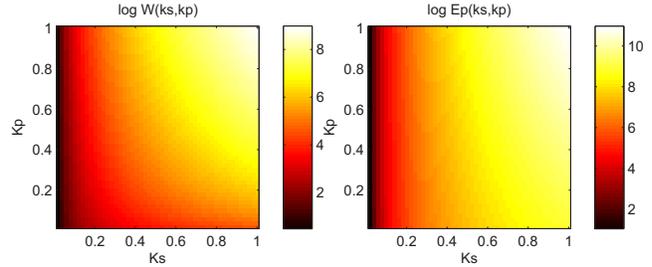}
\end{center}
\caption{Surface graph of the generated W and P-mode amplitudes at
$ky=2H^{-1}$, $kx=-60H^{-1}$, and different values of $k_P$ and
$k_S$. Initial perturbations are normalized to set E(0)=1.
Excitation amplitudes of the entropy perturbations show predominant
dependence on the $k_S$ (left panel), while only pressure
stratification scale $k_P$ is important for the generation of
P-modes (right panel). See electronic edition of the journal for
color images.} \label{surf_s}
\end{figure}

Fig. \ref{surf_s} shows the dependence of the W and P-mode
generation on the pressure and entropy stratification scales. As
expected from qualitative estimates, P-mode excitation depends
almost solely on the pressure stratification scale $k_P$, while the
generation of potential vorticity generally grows with $\eta$.

\subsection{P-mode: direct coupling with S-mode and indirect
coupling with W-mode}

\begin{figure}
\begin{center}
\includegraphics[width=84mm]{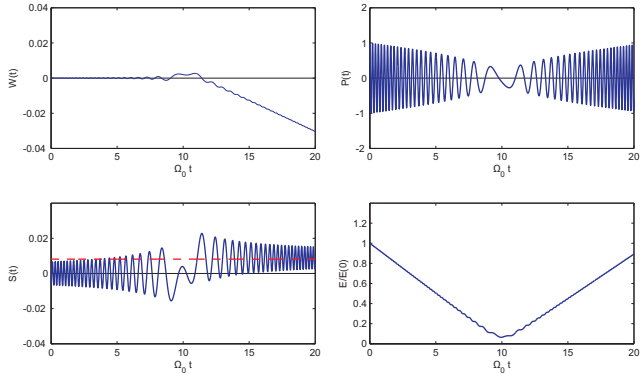}
\end{center}
\caption{Evolution of the P-mode SFH in the flow with
$k_x(t)=-30H^{-1}$, $k_y=2H^{-1}$ and equilibrium with growing
baroclinic perturbations $k_P=k_S=0.2H^{-1}$. Mode coupling occurs
in the vicinity of $t=10\Omega_0^{-1}$, where W and S-modes are
excited. The amplitude of the generated aperiodic contribution to
the entropy perturbation is marked by the red doted line. Farther,
this component leads to the baroclinic production of potential
vorticity with negative sign.} \label{SFH_p1}
\end{figure}

Fig. \ref{SFH_p1} shows the evolution of an initially imposed P-mode
SFH in a flow with growing baroclinic perturbations.  The {\it
oscillating} behavior of the entropy perturbation for $t < 10$ is
given by the P-mode. This oscillating component has a zero mean
value when averaged over time-scales longer than the wave period.
The existence of the {\it aperiodic} S-mode is instead characterized
by a nonzero mean value.  When the azimuthal wavenumber $k_y(t)$
changes sign at $t = 10$, we can observe the appearance of a nonzero
mean value (marked on the plot by the horizontal dashed line),
indicating that the high frequency oscillations of the P-mode are
able to generate the aperiodic perturbations of the S-mode.  The
aperiodic part of the entropy perturbation is than able to generate
potential vorticity perturbations. However, as we see from Eq. (54)
and Fig. \ref{coupling}, there is no direct coupling between P and
W-modes. Therefore, the P-mode generates the S-mode by shear flow
induced mode conversion, while the W-mode is further generated
because of its baroclinic ties with the entropy SFH. We describe
this situation as the three-mode coupling or in other words,
indirect coupling of the P to the W-mode. Note, that although the S
and W-mode generation is apparent from the dynamics of entropy and
potential vorticity SFH, energetically it plays a minor role as
compared to the compressible energy carried by the P-mode.

\begin{figure}
\begin{center}
\includegraphics[width=84mm]{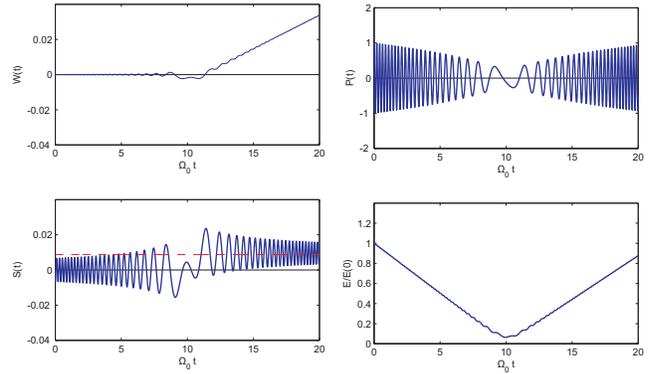}
\end{center}
\caption{Same as in previous figure but for $k_P=-0.2H^{-1}$ and $
k_S=0.2H^{-1}$. Perturbations are stable to baroclinic forces.
However, production of the potential vorticity with positive sign is
still observed.} \label{SFH_p2}
\end{figure}

Fig \ref{SFH_p2}. shows that P-mode generates potential vorticity
with a positive sign. However, the sign of the generated potential
vorticity depends on the initial phase of the P-mode. Hence, our
numerical results show generation of the W-mode with both positive
and negative signs.

It is interesting also to look at the P-mode dynamics in flows
stable to baroclinic perturbations (see Fig. \ref{SFH_p2}). The
initially imposed P-mode is able to generate the S-mode and
consequently the W-mode, that gives a growth of the potential
vorticity with time. Apart from the intrinsic limitations (the
dependence of the sign of the generated potential vorticity on the
initial phase of the P-mode and the low efficiency of the W-mode
generation), this process demonstrates the fact that potential
vorticity can be actually generated in flows with positive radial
buoyancy ($\eta<0$) and Richardson number.

\begin{figure}
\begin{center}
\includegraphics[width=84mm]{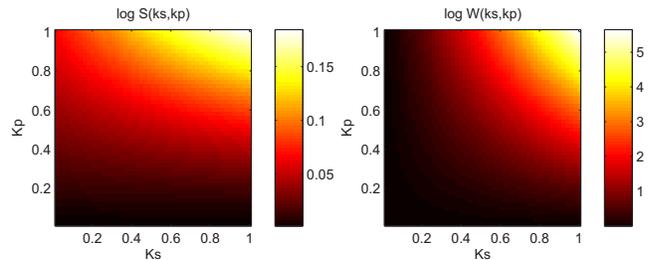}
\end{center}
\caption{Surface graph of the generated S and W-mode amplitudes at
$ky=2H^{-1}$, $kx=-60H^{-1}$, and different values of $k_P$ and
$k_S$. Initial perturbations are normalized to set E(0)=1.
Excitation amplitudes of the entropy perturbations mainly depend on
the $k_S$ (left panel), while both pressure and entropy
stratification scales are important for the generation of W-mode
perturbations (right panel). See electronic edition of the journal
for color images.} \label{surf_p}
\end{figure}

Fig. \ref{surf_p} shows the dependence of the S and W-mode
generation on the pressure and entropy stratification scales. In
good agreement with qualitative estimates, the S-mode excitation
depends strongly on the entropy stratification scale $k_S$, while
the generation of the potential vorticity generally grows with
$\eta$.

\section{Conclusion and Discussion}

We have studied the dynamics of linear perturbations in a 2D,
radially stratified, compressible, differentially rotating flow with
different radial density, pressure and entropy gradients. We
employed global radial scaling of linear perturbations and removed
the algebraic modulation due to the background stratification. We
derived a local dispersion equation for nonaxisymmetric
perturbations and the corresponding eigenfunctions in the zero shear
limit. We show that the local  stability of baroclinic perturbations
in the barotropic equilibrium state is defined by the
Schwarzschild-Ledoux criterion.

We study the shear flow induced linear coupling and the related
possibility of the energy transfer between the different modes of
perturbations using qualitative and a more detailed numerical
analysis. We employ a three-mode formalism and describe the behavior
of S W and P-modes under the action of the baroclinic and velocity
shear forces in local approximation.

We find that the system exhibits an asymmetric coupling pattern with
five energy exchange channels between three different modes. The
W-mode is coupled to S and P-modes: perturbations of the potential
vorticity are able to excite entropy and compressible modes. The
amplitude of the generated S-mode grows with the increase of entropy
stratification scale of the background ($k_S$) while the amplitude
of the generated P-mode perturbations grows with the increase of
background baroclinic index ($\eta$). The S-mode is coupled to the W
and P-modes: the amplitude of the generated P-mode perturbations
grows with increase of the background pressure stratification scale
($k_P$), while the amplitude of the W-mode grows with the increase
of baroclinic index. The P-mode is coupled to the S-mode: the
amplitude of generated entropy perturbations grows with the increase
of the background entropy stratification scale. On the other hand,
there is no direct energy exchange channel from P to W mode and,
therefore, no direct conversion is possible. Our results, however,
show that the P-mode is still able to generate W-mode through
indirect three mode P-S-W coupling scheme. This linear inviscid
mechanism indicates that compressible perturbations are able to
generate potential vorticity via aperiodic entropy perturbations.

The dynamics of radially stratified discs have been already studied
by both, linear shearing sheet formalism and direct numerical
simulations. However, previous studies focus on the baroclinic
stability and vortex production by entropy perturbations, neglecting
the coupling with higher frequency density waves.

The most vivid signature of  density wave excitation in radially
stratified disc flows can be seen in Klahr (2004). The numerical
results presented on the linear dynamics of perturbation SFH show
high frequency oscillations after the radial wavenumber changes
sign. However, focusing on the energy dynamics, the author filters
out high frequency oscillations from the analysis.

The purpose of numerical simulations by Johnson and Gammie (2006)
was the investigation of the velocity shear effects on the radial
convective stability and the possibility of the development of
baroclinic instability. Therefore, no significant amount of
compressible perturbations is present initially, and it is hard to
judge if high frequency oscillations appear later in simulations.
Petersen et al. (2007a), (2007b) employed the anelastic
approximation that does not resolve the coupling of potential
vorticity and entropy with density waves. Moreover, if produced,
high frequency density waves soon develop into  spiral shocks (see
e.g., Bodo et al 2007). The anelastic gas approximation does
intentionally neglect this complication and simplifies the
description down to low frequency dynamics.

Numerical simulations of hydrodynamic turbulence in unstratified
disc flows showed that the dominant part of  turbulent energy is
accumulated into the high frequency compressional waves (see, e.g.,
Shen et al. 2006). On the other hand, it is vortices that are
thought to play a key role in hydrodynamic turbulence in accretion
discs, as well as planet formation in protoplanetary disc dynamics.
Therefore, any link and possible energy exchange between high
frequency compressible oscillations and aperiodic vortices can be an
important factor in the above described astrophysical situations.

Based on the present findings we speculate that density waves can
participate in the process of the development of regular vortical
structures in discs with negative radial entropy gradients.
Numerical simulations have shown that thermal (entropy)
perturbations can generate vortices in baroclinic disc flows (see
e.g., Petersen et al. 2007a, 2007b). Hence, vortex development
through this mechanism depends on the existence of initial regular
entropy perturbations, i.e., thermal plumes, in differentially
rotating baroclinic disc flows.

It seems that compressional waves with linear amplitudes can heat
the flow through two different channels: viscous dissipation and
shear flow induced mode conversion. However, there is a strict
difference between the entropy production by the kinematic shear
mechanism and viscous dissipation. In the latter case, compressional
waves first need to be tightly stretched down to the dissipation
length-scales by the background differential rotation to be  subject
of viscous dumping. As a result, the entropy produced by viscous
dissipation of compressional waves takes a shape of narrow stretched
lines. This thermal perturbations can baroclinically produce
potential vorticity of similar configuration. However, this is
clearly not an optimal form of potential vorticity that can lead to
the development of the long-lived vortical structures. On the
contrary, entropy perturbations produced through the mode conversion
channel can have a form of a localized thermal plumes. These can be
very similar to those used in numerical simulations by Petersen et
al. (2007a), (2007b). In this case compressional waves can
eventually lead to the development of  persistent vortical
structures of different polarity. Hence, high frequency oscillations
of the P mode can participate in the generation of anticyclonic
vortices that further accelerate dust trapping and planetesimal
formation in protoplanetary discs with equilibrium entropy
decreasing radially outwards.

Using the local linear approximation we have shown the possibility
of the potential vorticity generation in flows with both, positive
and negative radial entropy gradients (Richardson numbers). In fact,
the standard alpha description of the accretion discs implies
\emph{positive} radial stratification of entropy and hence, weak
baroclinic decay of existing vortices. In this case there will be a
competition between the ``baroclinic viscosity'' and potential
vorticity generation due to  mode conversion. Hence, it is not
strictly overruled that a significant amount of compressional
perturbations can lead to the development of anticyclonic vortices
even in flows with positive entropy gradients. In this case, radial
stratification opens an additional degree of freedom for velocity
shear induced mode conversion to operate. Although, the viability of
this scenario needs further investigation.

This paper presents the results obtained within the linear shearing
sheet approximation. At nonlinear amplitudes, the P mode leads to
the development of shock waves. These shocks induce local heating in
the flow. Therefore, a realistic picture of entropy production and
vortex development in radially stratified discs with significant
amount of compressible perturbations needs to be analyzed by direct
numerical simulations.

\section*{Acknowledgments}

A.G.T. was supported by GNSF/PRES-07/153. A.G.T. would like to
acknowledge the hospitality of Osservatorio Astronomico di Torino.
This work is supported in part by ISTC grant G-1217.

{}

\appendix

\section{Initial conditions}

Here we present the approximations used to derive the analytic form
of the initial conditions corresponding to individual modes in
radially stratified shear flows. These conditions are used to
construct the initial values of perturbations in the numerical
integration of the ODEs governing the linear dynamics of
perturbations in these flows. We employ different methods for high
and low frequency modes.

\subsection{P-mode}

P-mode perturbations are intrinsically high frequency and well
separated from low frequency modes everywhere outside the coupling
region $k_x/k_y < 1$. In order to construct P-mode perturbations we
use convective eigenfunction derived in the shearless limit and
account for shear flow effects only in the adiabatic limit:
\begin{equation}
\Psi_{c}(t) = (\omega_{c}^2(t) + c_s^2 \eta) P(t) - 2 \Omega_0 W(t)
- c_s^2 k_P k_x(t) s(t) ~,
\end{equation}
where
\begin{equation}
\omega_c^2(t) = -{c_s^2 \eta k_y^2 \over c_s^2 k^2(t) - 4B\Omega_0}
~.
\end{equation}
Although this form of the eigenfunction is not valid function for
describing W and S modes individually in a sheared medium, it has
proved to be a good tool for excluding both modes from the initial
spectrum:
\begin{equation}
\Psi_{c}(0) = 0 ~.
\end{equation}
Assuming that we are looking for P-mode perturbations with
wave-numbers satisfying the condition $ k_x(0)/k_y \gg 1 $ we may
use the zero potential vorticity condition:
\begin{equation}
W(0) = 0 ~.
\end{equation}
Hence, Eqs. (A3,A4) yield the full set of initial conditions for the
high frequency P-mode SFH of perturbations:
\begin{equation}
p(0) = P_0 ~, ~~~~~ u_x(0) = U_0 ~,
\end{equation}
\begin{equation}
u_y(0) = {1 \over k_x(0)} \left( k_y U_0 + 2BP_0 \right) ~,
\end{equation}
\begin{equation}
s(0) = {\omega_c^2(0)+c_s^2 \eta \over c_s^2 k_p k_x(0) } P_0 ~,
\end{equation}
where $P_0$ and $U_0$ are free parameters corresponding to the two
P-modes in the system. Specific values of these two parameters
define whether the  potential or kinetic part of the wave harmonic
is present initially.

\subsection{Low frequency modes}

In order to derive the initial conditions for the S and W modes
individually we employ the second order equation for radial velocity
perturbation that can be derived from Eqs. (34-37):
\begin{equation}
\left\{ {{\rm d}^2 \over {\rm d} t^2} + c_s^2 k^2 - 4B\Omega_0 -
c_s^2 \eta \right\} u_x = -c_s^2 k_y W + 4Ac_s^2 k_y p ~.
\end{equation}
\begin{equation}
\left\{ {{\rm d}^2 \over {\rm d} t^2} + c_s^2 k^2 - 4B\Omega_0
\right\} u_y = c_s^2 k_x(t) W + 2 B c_s^2 k_P s ~,
\end{equation}
For low frequency perturbations
\begin{equation}
{{\rm d}^2 \over {\rm d} t^2} \left( \begin{array}{c} u_x \\ u_y
\end{array} \right) \sim \omega^2_c \left( \begin{array}{c} u_x \\ u_y
\end{array} \right) ~.
\end{equation}
Assuming that $\omega_c^2(0) \ll c_s^2 k^2(0)$ and neglecting the
corresponding terms in Eqs. (A6-A7) leads to the following algebraic
system:
\begin{equation}
\left[c_s^2 k^2 - 4B\Omega_0 \right] u_x = -c_s^2 k_y W + 4Ac_s^2
k_y p ~.
\end{equation}
\begin{equation}
\left[c_s^2 k^2 - 4B\Omega_0 \right] u_y = c_s^2 k_x(t) W + 2 B
c_s^2 k_P s ~.
\end{equation}
Hence, we can derive the initial conditions for the low frequency
modes as follows:
\begin{equation}
p(0) = {B \over 2A c_s^2 k_y^2 + B\omega_p^2(0)} \left( 2 \Omega_0
W_0 + c_s^2 k_p k_x(0) S_0 \right) ~,
\end{equation}
\begin{equation}
u_x(0) = {1 \over \omega_p^2(0)} \left( -c_s^2 k_y W_0 + 4A c_s^2
k_y p(0) \right) ~,
\end{equation}
\begin{equation}
u_y(0) = {1 \over \omega_p^2(0)} \left( c_s^2 k_x(0) W_0 + 2B c_s^2
k_p S_0 \right) ~,
\end{equation}
where
\begin{equation}
\omega_p^2(0) = c_s^2 (k_x^2(0) + k_y^2) - 4B\Omega_0 ~.
\end{equation}
Eqs. (A11-A14) give the initial values of perturbation SFHs for
S-mode when
\begin{equation}
W_0 = 0 ~,~~~ S_0 \not= 0 ~,
\end{equation}
and W-mode when
\begin{equation}
W_0 \not= 0 ~,~~~ S_0 = 0 ~.
\end{equation}

\end{document}